\documentclass[12pt]{article}

\usepackage{verbatim}
\usepackage{amsmath}
\usepackage{amssymb}
\usepackage{graphicx}
\usepackage{cite}
\usepackage{subfig}
\usepackage{setspace}
\usepackage{url}
\usepackage[percent]{overpic}
\usepackage{slashed}
\usepackage{authblk}

\usepackage{xspace}

\usepackage{fullpage}
\usepackage{hyperref}
\def\tev{\,{\rm TeV}}
\def\gev{\,{\rm GeV}}
\def\ie{{\it i.e.}}
\def\eg{{\it e.g.}}

\def\etal{{\it et al.}}

\def\to{\rightarrow}

\title{Constraints on Higgs Properties and SUSY Partners in the pMSSM{\footnote {White Paper contributed to the Snowmass Community Summer Study 2013, Minneapolis, MN July 29 - August 6, 2013}}}
\date{September 26, 2013}
\author{M. Cahill-Rowley}
\author{J. Hewett}
\author{A. Ismail}
\author{T. Rizzo}
\affil{SLAC National Accelerator Laboratory, Menlo Park, CA, USA\footnote{mrowley, hewett, aismail, rizzo@slac.stanford.edu}}

\begin{document}

\rightline{\vbox{\halign{&#\hfil\cr
&SLAC-PUB-15561\cr
}}}


{\let\newpage\relax\maketitle}

\begin{abstract}
Direct searches for superpartners and precision measurements of the properties of the $\sim 126$ GeV Higgs boson lead to important inter-dependent 
constraints on the underlying parameter space of the MSSM.  The 19/20-parameter p(henomenological)MSSM offers a flexible framework for the study 
of a wide variety of both Higgs and SUSY phenomena at the LHC and elsewhere. Within this scenario we address the following questions: `What will potentially null 
searches for SUSY at the LHC tell us about the possible properties of the Higgs boson?' and, conversely, `What do precision measurements of the properties 
of the Higgs tell us about the possible properties of the various superpartners?' Clearly the answers to such questions will be functions of both the 
collision energy of the LHC as well as the accumulated integrated luminosity. We address these questions employing several sets of pMSSM models having either 
neutralino or gravitino LSPs, making use of the ATLAS SUSY analyses at the 7/8 TeV LHC as well as planned SUSY and Higgs analyses at the 14 TeV LHC and the ILC. Except for theoretical uncertainties 
that remain to be accounted for in the ratios of SUSY and SM couplings, we demonstrate that Higgs coupling measurements at the 14 TeV LHC, and particularly at the 500 GeV ILC, 
will be sensitive to regions of the pMSSM model space that are not accessible to direct SUSY searches. 
\end{abstract}

\section{Introduction and Overview of the pMSSM Models}

With the discovery of the Higgs boson, the last component of the Standard Model is now in place and searches for new physics continue 
in earnest. A most important issue with respect to the production of new physics at an accelerator is whether or not it can be discovered or 
excluded given backgrounds arising from the Standard Model (SM), provided it is kinematically
accessible. In particular, within a specific model, we would like to know 
how well a given set of experimental analyses can probe the full parameter space of interest. With the lack of any experimental 
evidence for new physics so far, this is certainly true in the case of Supersymmetry (SUSY). However, even in the simplest SUSY scenario, the MSSM, 
the number of free parameters ($\sim$ 100) is too large to study in complete generality. The traditional approach 
is to assume the existence of some high-scale theory with only a few parameters (such as mSUGRA{\cite {SUSYrefs}) from which all 
the properties of the sparticles at the TeV scale can be determined and studied in detail. While such an approach is often quite 
valuable~\cite{Cohen:2013kna}, these scenarios are phenomenologically limiting and are under increasing tension with a 
wide range of experimental data including the $\sim 126$ GeV mass of the recently discovered Higgs boson{\cite {ATLASH,CMSH}}.  
Of course the discovery of the Higgs boson itself might also provide a new and important window into whatever physics lies beyond the SM. This 
is certainly true in the case of the MSSM where all of the properties of the Higgs can in principle be calculated perturbatively  
from the assumed values of the soft breaking parameters in the underlying Lagrangian.

One way to circumvent such limitations is to examine the more general 19/20-parameter pMSSM{\cite{Djouadi:1998di}}. The 
increased dimensionality of the parameter space not only allows for a more unprejudiced study of SUSY, but can also yield valuable
information on `unusual' scenarios, identify weaknesses in the current LHC analyses, and can be used to combine results from
many independent SUSY related searches. To these ends, we have recently embarked on a detailed study of the signatures for the pMSSM at the 7 and 8 TeV LHC, 
supplemented by input from Dark Matter (DM) experiments as well as from precision electroweak and flavor 
measurements{\cite {us1,us2}}. The pMSSM is the most general version of the R-parity conserving MSSM when it is subjected to  
a minimal set of experimentally-motivated guiding principles: ($i$) CP conservation, ($ii$) Minimal Flavor Violation at the 
electroweak scale so that flavor physics 
is controlled by the CKM mixing matrix, ($iii$) degenerate 1\textsuperscript{st} and 2\textsuperscript{nd} generation sfermion masses,
and ($iv$) negligible Yukawa couplings and A-terms for the first two generations. 
In particular, no assumptions are made about physics at high scales, e.g., the nature of SUSY 
breaking, in order to capture electroweak scale phenomenology for which a UV-complete theory may not yet exist. Imposing these principles  decreases the number of free parameters in the MSSM at the TeV-scale from 105 to 19 for the case of a neutralino LSP, or to 
20 when the gravitino mass is included as an additional parameter when it plays the role of the LSP.  We have not assumed that the LSP relic density 
necessarily saturates the WMAP/Planck value{\cite{Komatsu:2010fb}} in order to allow for the possibility of multi-component 
DM. For example, the axions introduced to solve the strong CP problem may contribute to the DM relic density. The 19/20 pMSSM parameters 
and the ranges of values employed in our scans are listed in Table~\ref{ScanRanges}. Like throwing darts, to study the pMSSM   
we generate $\sim 3.7 \times 10^6$ model points in this space (using SOFTSUSY{\cite{Allanach:2001kg}} and checking for consistency with 
SuSpect{\cite{Djouadi:2002ze}}), with each point then corresponding to a specific set of values for these parameters. These individual models 
are then subjected to a large set of collider, flavor, precision measurement, dark matter and theoretical constraints~\cite{us1}.  
Roughly $\sim$225k models for each type of LSP survive this initial selection and can then be used for further physics studies. Decay 
patterns of the SUSY partners and the extended Higgs sector are calculated using a privately modified version of SUSY-HIT{\cite{Djouadi:2006bz}} 
as well as the most recent version of HDECAY{\cite {HDECAY}}. 
Since our scan ranges include sparticle masses up to 4 TeV, an upper limit chosen to enable phenomenological studies at the 14 TeV LHC, 
the neutralinos and charginos in either of the model sets are typically very close to being a pure 
electroweak eigenstate as the off-diagonal elements of the corresponding mass matrices are at most $M_W$. This has important 
implications for the resulting collider and DM phenomenology\cite{Cahill-Rowley:2013yla}. Finally, for the neutralino (gravitino) model set we find that roughly 
$\simeq 20 (10)\%$ of the models are found to satisfy $m_h=126 \pm 3$ GeV; we will focus on these subsets in the analysis that follows\footnote{We note that our model sets were generated before the Higgs boson was discovered.}.

In addition to these two large pMSSM model sets, we have also generated a smaller, specialized neutralino LSP set of $\sim$ 10.2k 
`natural' models, all of which predict $m_h=126\pm 3$ GeV, have an LSP that {\it does} saturate the WMAP relic density and produce 
values of fine-tuning (FT) better than $1\%$ using the Ellis-Barbieri-Giudice measure~\cite{Ellis:1986yg, Barbieri:1987fn}. This low-FT model 
set will also be used as part of the present study. In order to obtain this model set we modified the parameter scan ranges listed in the Table to 
greatly increase the likelihood of both low FT and having a thermal relic density in the desired range.  Amongst other things, this requires 
a bino as the LSP, as well as light Higgsinos and highly mixed stops. We generated $\sim 3.3 \times 10^8$ low-FT points 
in this 19-parameter space and subjected them to updated precision, flavor, DM and collider constraints as before. Since these requirements 
were much stricter than for our two larger model sets, only $\sim$ 10.2k low-FT models were found to be viable for further study. 

Within each pMSSM model, the properties of the lightest CP-even Higgs, $h$, are completely determined as are 
the corresponding properties of all the superpartners since these follow directly from the chosen values of the soft parameters.  This allows us to address 
the two questions: What will potentially null searches for SUSY at the LHC tell us about the possible properties of the Higgs Boson, and
what do precision measurements of the properties of the Higgs Boson tell us about the possible properties of the superpartners?

\begin{table}
\centering
\begin{tabular}{|c|c|} \hline\hline
$m_{\tilde L(e)_{1,2,3}}$ & $100 \gev - 4 \tev$ \\ 
$m_{\tilde Q(q)_{1,2}}$ & $400 \gev - 4 \tev$ \\ 
$m_{\tilde Q(q)_{3}}$ &  $200 \gev - 4 \tev$ \\
$|M_1|$ & $50 \gev - 4 \tev$ \\
$|M_2|$ & $100 \gev - 4 \tev$ \\
$|\mu|$ & $100 \gev - 4 \tev$ \\ 
$M_3$ & $400 \gev - 4 \tev$ \\ 
$|A_{t,b,\tau}|$ & $0 \gev - 4 \tev$ \\ 
$M_A$ & $100 \gev - 4 \tev$ \\ 
$\tan \beta$ & 1 - 60 \\
$m_{3/2}$ & 1 eV$ - 1 \tev$ ($\tilde{G}$ LSP)\\
\hline\hline
\end{tabular}
\caption{Scan ranges for the 19 (20) parameters of the pMSSM with a neutralino (gravitino) LSP. The gravitino mass is scanned with 
a log prior. All other parameters are scanned with flat priors, though we expect this choice to have little qualitative impact on 
our results based on previous studies~\cite{Djouadi:1998di,us}.}
\label{ScanRanges}
\end{table}

\section{LHC SUSY Searches}

First, we must determine how well searches at the 7, 8 and eventually 
$\sim$14 TeV LHC will probe the pMSSM parameter sets that we have generated. Once these constraints on the pMSSM space are known, we can 
determine how the properties of the lightest Higgs may differ from those of the SM in light of the (so far) 
null SUSY search results. Correlations between the direct search results and the properties of the Higgs then provide an answer to the first question that 
we posed above. 

To this end we begin this step of the analysis with a short overview of the searches for the pMSSM at the 7 and 8 TeV LHC; the same overall 
approach will carry over to our 
14 TeV study. In general, we follow the suite of ATLAS SUSY search analyses as closely as possible employing fast Monte Carlo. These are also 
supplemented by several searches performed by CMS. The specific analyses applied to the neutralino model set are briefly summarized in 
Tables~\ref{SearchList7} and~\ref{SearchList8}. We augment the standard SUSY searches by including searches for heavy stable charged particles and heavy neutral 
SUSY Higgs decay into $\tau^+\tau^-$ as performed by CMS~\cite{CMSextra} and measurements of the rare decay mode $B_s\to \mu^+\mu^-$ as discovered by 
CMS and LHCb~\cite{BSMUMU}.  All of these play distinct but important roles in restricting the pMSSM parameter space. Presently, we have 
implemented every relevant ATLAS SUSY search publicly available as of the beginning of March 2013. This list is currently being 
updated and expanded for future analysis. The analysis results for all three of these model sets discussed here appear in detail in our companion HE4 Snowmasss 
White Paper on SUSY searches~\cite{Cahill-Rowley:2013yla}.

\begin{table}
\centering
\begin{tabular}{|l|l|c|c|c|} \hline\hline
Search & Reference & Neutralino & Gravitino & Low-FT   \\
\hline
2-6 jets & ATLAS-CONF-2012-033  & 21.2\% &  17.4\% & 36.5\% \\
multijets & ATLAS-CONF-2012-037 & 1.6\%  & 2.1\% & 10.6\% \\
1-lepton & ATLAS-CONF-2012-041 & 3.2\%  & 5.3\% & 18.7\%  \\

HSCP      &  1205.0272  & 4.0\% & 17.4\% & $<$0.1\%  \\
Disappearing Track  & ATLAS-CONF-2012-111 & 2.6\%  & 1.2\% & $<$0.1\% \\
Muon + Displaced Vertex  & 1210.7451 & - & 0.5\% & - \\
Displaced Dilepton & 1211.2472 & - & 1.1\% & - \\

Gluino $\to$ Stop/Sbottom   & 1207.4686 & 4.9\% &  3.5\% & 21.2\% \\
Very Light Stop  & ATLAS-CONF-2012-059 & $<$0.1\% & $<$0.1\% & 0.1\%  \\
Medium Stop  & ATLAS-CONF-2012-071 & 0.3\% & 5.1\% & 2.1\% \\
Heavy Stop (0l)  & 1208.1447 & 3.7\% & 3.0\% & 17.0\% \\
Heavy Stop (1l)   & 1208.2590 & 2.0\% & 2.2\% & 12.6\% \\
GMSB Direct Stop  & 1204.6736 & $<$0.1\% & $<$0.1\% & 0.7\% \\
Direct Sbottom & ATLAS-CONF-2012-106 & 2.5\% & 2.3\% & 5.1\% \\
3 leptons & ATLAS-CONF-2012-108 & 1.1\% & 6.1\% & 17.6\% \\
1-2 leptons & 1208.4688 & 4.1\% & 8.2\% & 21.0\% \\
Direct slepton/gaugino (2l)  & 1208.2884 & 0.1\% & 1.2\% & 0.8\% \\
Direct gaugino (3l) & 1208.3144 & 0.4\% & 5.4\% & 7.5\% \\
4 leptons & 1210.4457 & 0.7\% & 6.3\% & 14.8\% \\
1 lepton + many jets & ATLAS-CONF-2012-140 & 1.3\% & 2.0\% & 11.7\% \\
1 lepton + $\gamma$ & ATLAS-CONF-2012-144 & $<$0.1\% & 1.6\% & $<$0.1\% \\
$\gamma$ + b & 1211.1167 & $<$0.1\% & 2.3\% & $<$0.1\% \\
$\gamma \gamma $ + MET & 1209.0753 & $<$0.1\% & 5.4\% & $<$0.1\% \\

$B_s \to \mu \mu$ & 1211.2674 & 0.8\% & 3.1\% & * \\
$A/H \to \tau \tau$ & CMS-PAS-HIG-12-050 & 1.6\% & $<$0.1\% & * \\

\hline\hline
\end{tabular}
\caption{7 TeV LHC searches included in the present analysis and the corresponding fraction of the neutralino, gravitino and low-FT pMSSM 
model sets excluded by each search. Note that in the case of the last two entries the experimental constraints have already been included 
in the model generation process for the low-FT model set and therefore are not shown here.}
\label{SearchList7}
\end{table}

\begin{table}
\centering
\begin{tabular}{|l|l|c|c|c|} \hline\hline
Search & Reference & Neutralino & Gravitino & Low-FT    \\
\hline

2-6 jets   & ATLAS-CONF-2012-109 & 26.7\% & 21.6\% & 44.9\% \\
multijets   & ATLAS-CONF-2012-103 & 3.3\% & 3.8\% & 20.9\% \\
1-lepton     & ATLAS-CONF-2012-104 & 3.3\% & 6.0\% & 20.9\% \\
SS dileptons & ATLAS-CONF-2012-105 & 4.9\% & 12.4\% & 35.5\% \\

Medium Stop (2l) & ATLAS-CONF-2012-167 & 0.6\% & 8.1\% & 4.9\% \\
Medium/Heavy Stop (1l) & ATLAS-CONF-2012-166 & 3.8\% & 4.5\% & 21.0\% \\
Direct Sbottom (2b) & ATLAS-CONF-2012-165 & 6.2\% & 5.1\% & 12.1\% \\
3rd Generation Squarks (3b) & ATLAS-CONF-2012-145 & 10.8\% & 9.9\% & 40.8\% \\
3rd Generation Squarks (3l) & ATLAS-CONF-2012-151 & 1.9\% & 9.2\% & 26.5\% \\
3 leptons & ATLAS-CONF-2012-154 & 1.4\% & 8.8\% &32.3\% \\
4 leptons & ATLAS-CONF-2012-153 & 3.0\% & 13.2\% & 46.9\% \\
Z + jets + MET & ATLAS-CONF-2012-152 & 0.3\% & 1.4\% &6.8\% \\

\hline\hline
\end{tabular}
\caption{Same as in the previous table but now for the 8 TeV ATLAS MET-based SUSY searches. Note that when all the above searches in both Tables are combined for the neutralino 
(gravitino, low-FT) model set we find that $\sim 37~(52,~70)\%$ of these models are currently excluded by the LHC.}
\label{SearchList8}
\end{table}

Very briefly stated, our procedure is as follows: We generate SUSY events for each model for all relevant (up to 85) production channels in PYTHIA 
6.4.26~\cite{Sjostrand:2006za}, and then pass the events through fast detector simulation using PGS 4~\cite{PGS}. Both programs have been modified to, 
e.g., correctly deal with gravitinos, multi-body decays, hadronization of stable colored sparticles, and ATLAS b-tagging. We then scale our 
event rates to NLO by calculating the relevant K-factors using Prospino 2.1~\cite{Beenakker:1996ch}. The individual searches are then 
implemented using our customized analysis code{\cite {us}}, which follows the published cuts and selection criteria as closely as possible. 
This analysis code is validated for each of the many search regions for every analysis employing the benchmark model points provided by ATLAS 
(and CMS). Models are then excluded using the 95\% $CL_s$ limits as obtained by ATLAS (and CMS). For the two large model sets these analyses are 
performed {\it without} requiring the Higgs mass constraint, $m_h=126 \pm 3$ GeV (combined experimental and theoretical errors) so that we 
can understand its influence on the search results. Recall that roughly $\sim 20 (10)\%$ of models in the neutralino (gravitino) model set 
predict a Higgs mass in the above range. While there is some variation amongst the individual searches themselves, we find that, 
once combined, the total fraction of our models surviving the set of all LHC searches is to an excellent approximation 
{\it independent} of whether or not the Higgs mass constraint has been applied. Conversely, the $\sim 20(10)\%$ fraction of 
neutralino (gravitino) models predicting the correct Higgs mass is also found to be approximately independent of whether  the SUSY searches have been applied.
This result is very powerful and demonstrates the approximate decoupling of direct SUSY search results from the discovery of the Higgs boson, allowing us, in general, to continue examining the properties and signatures of the entire neutralino and gravitino model samples with some reasonable validity. Of course, for this study, in which we examine the properties of the Higgs boson itself, we restrict our analysis to the subsets of the neutralino and gravitino model sets which predict $m_h=126 \pm 3$ GeV. No additional requirements on the Higgs mass are necessary for the low-FT set, since in this case the Higgs mass constraint is imposed during model generation.

\section{Determination of Higgs Properties in the pMSSM}

The next part of this project is to use the measured properties of the Higgs to directly constrain the pMSSM parameter space. To do this we determine 
the extent to which the couplings of the light CP-even Higgs boson in the pMSSM differ from the Standard Model expectations, then compare these results to current and future experimental determinations of 
the couplings. We make several comparisons corresponding to the anticipated evolution of our knowledge about 
the allowed values of the couplings: ($i$) limits from current data{\cite {now}}, ($ii$) limits that are expected to be attainable at the 14 TeV LHC with an integrated luminosity 
of 0.3(3) ab$^{-1}${\cite {14TeVLHC}}, and finally ($iii$) projected limits from the ILC{\cite {ILC}}. To this end we employ the latest version of HDECAY 5.11 to calculate 
these coupling ratios 
in the analysis that follows. Note that since the full SUSY loop corrections for the $h \to WW$ and $h\to ZZ$ partial widths are not yet incorporated in HDECAY, we 
unfortunately can not employ these very important modes to constrain our model sample. 

We will follow the standard approach here and define the signal strength for production of a final state produced from the decay $h \to X$ via a 
given production channel (e.g. $gg,VBF\to h$), normalized to the corresponding SM value as 
\begin{equation}
\mu_{gg,VBF}(X) = {{\sigma(gg,VV\to h)~B(h \to X)}\over {SM}}\,.
\end{equation}
For final states $X$ which do not involve the top quark, we can also define the ratio of the squares of the various effective couplings to 
their corresponding SM values by simply forming the ratios of the relevant partial decay widths,
\begin{equation}
r_X = {{\Gamma (h \to X)}\over {SM}}\,,
\end{equation}
for the final states $X=ZZ,~W^+W^-,~\bar b b, ~\bar c c, ~\tau^+\tau^-,~gg, ~\gamma\gamma, ~\gamma Z$; the case of the $ht\bar t$ coupling must be 
handled separately and can only be directly accessed via associated production. We are, of course, also interested in the branching fraction for the invisible Higgs decays into, 
\eg, the lightest neutralino, producing a final state which is pure MET{\footnote {Decays to other sparticles are kinematically forbidden as a result of the LEP limits on charged 
particles; neutral winos and Higgsinos as well as sneutrinos are required to have a charged partner with a similar mass, preventing them from being decay products of a 
$\sim 126$ GeV Higgs.}}. Searches for invisible decays into these LSPs are very interesting because of their potential to place significant 
constraints on the SUSY parameter space, particularly when  results from ILC500 are employed.

\begin{figure}[htbp]
\centerline{\includegraphics[width=3.5in]{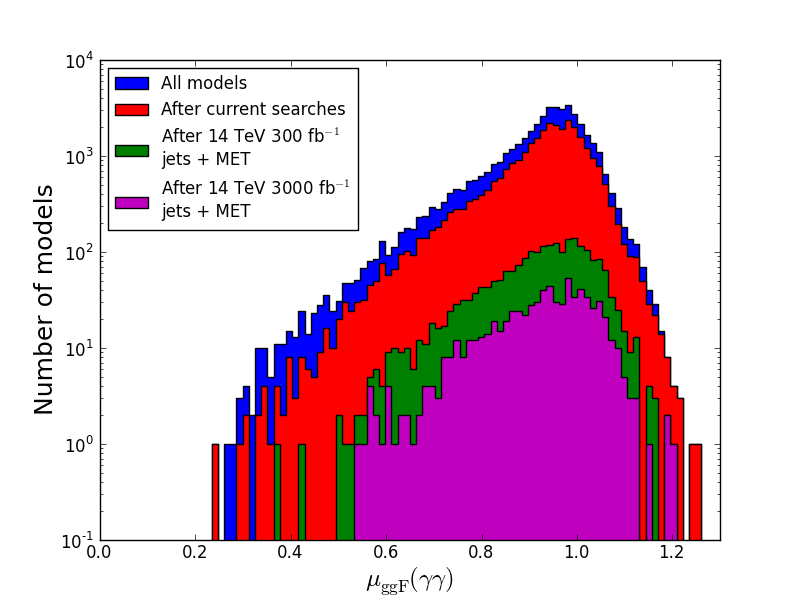}
\hspace{-0.50cm}
\includegraphics[width=3.5in]{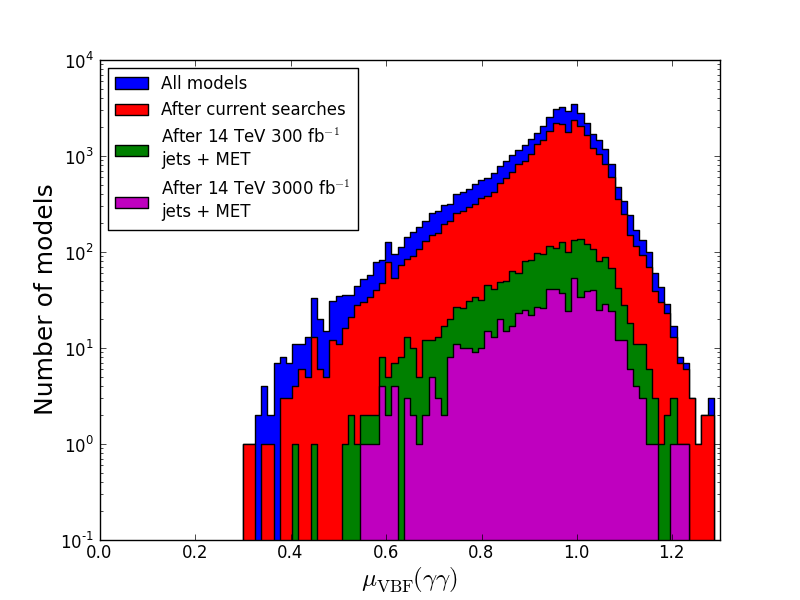}}
\vspace*{0.50cm}
\centerline{\includegraphics[width=3.5in]{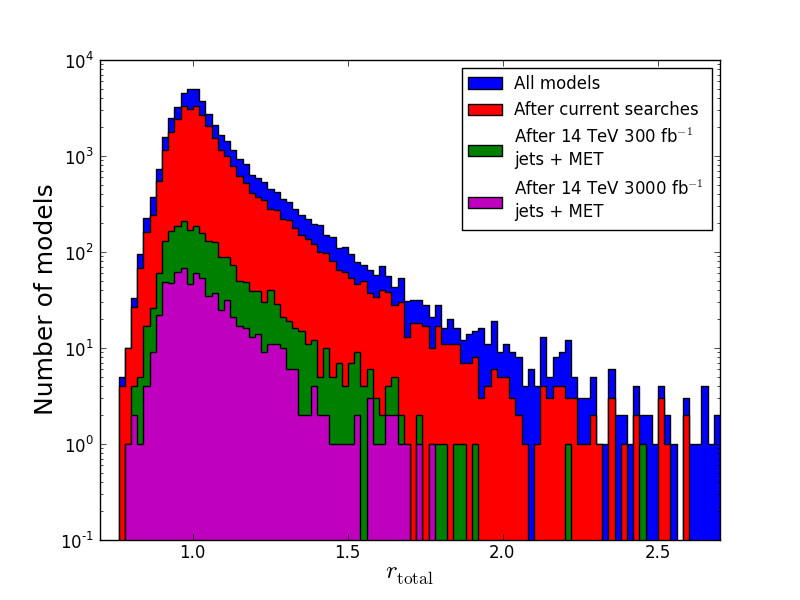}}
\vspace*{-0.10cm}
\caption{Histograms of signal strengths for $h \to \gamma \gamma$ in the $gg$-fusion (top left) and vector boson fusion (top right) channels for the subset of neutralino 
models that predict $m_h=126\pm 3$ GeV. The blue (red) histogram represents models before any ATLAS searches are applied (after the 7 and 8 TeV SUSY searches) 
while the green (purple) histograms show models that are expected to survive the zero-lepton jets plus MET search at 14 TeV, assuming a luminosity of 300 (3000) 
fb$^{-1}$. The ratio $r_{total}$ for the total $h$ width is analogously shown in the bottom panel.}
\label{figA}
\end{figure}

To get an initial idea of the distribution of Higgs properties in the various model sets it is instructive to look at a few examples. Consider 
Fig.~\ref{figA}, which shows the distribution of the $h\to \gamma\gamma$ signal strength for both the $gg$-fusion and vector boson fusion production channels in the neutralino 
LSP model subset with $m_h=126 \pm 3$ GeV, along with the effect of current and future ATLAS searches on this distribution~\cite{Cahill-Rowley:2013yla}. Other than the obvious fact that these distributions 
peak near unity but have long tails, the most important thing to notice is that the {\it shape} of these distributions (up to statistical fluctuations) is essentially unaffected 
by the ATLAS direct SUSY searches. Furthermore, the shape of the $r_{total}= \Gamma(h\to All)/SM$ distribution for the neutralino set also shows that this shape invariance 
is maintained for the other observables.  We therefore see that SUSY searches and Higgs boson properties are to a very good approximation 
orthogonal. As we will show below, the other final states exhibit a similar behavior, answering the first question posed in the abstract - future null searches at the LHC will, to a good approximation, not affect the range of values that we might expect for the Higgs couplings.

\begin{figure}[htbp]
\centerline{\includegraphics[width=5.5in]{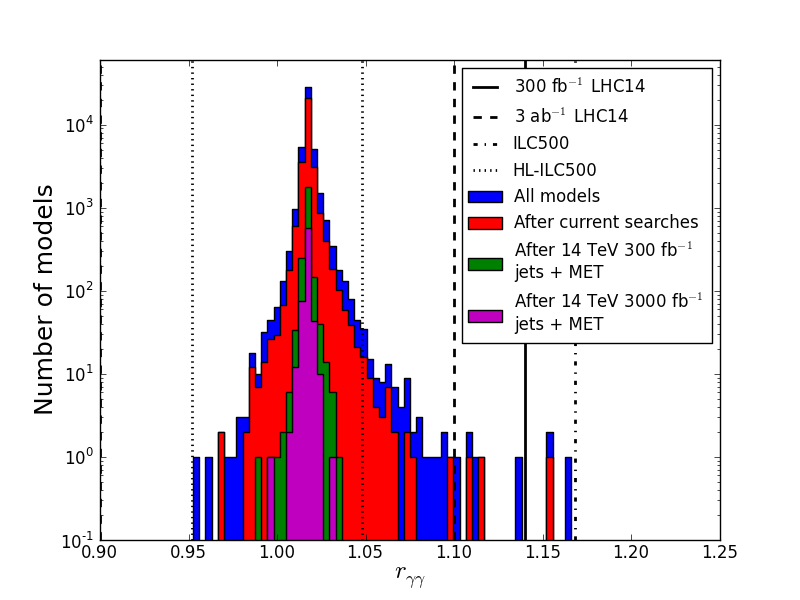}}
\vspace*{0.50cm}
\centerline{\includegraphics[width=3.5in]{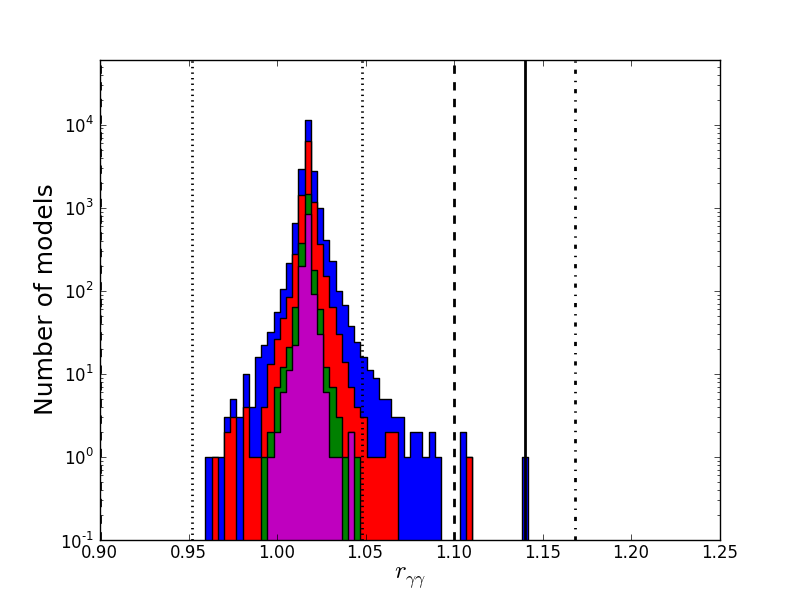}
\hspace{-0.50cm}
\includegraphics[width=3.5in]{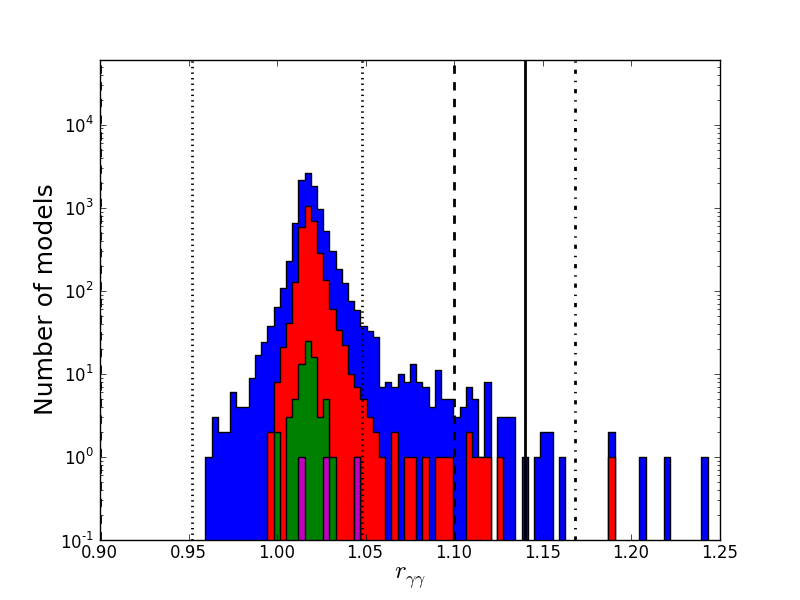}}
\vspace*{-0.10cm}
\caption{Histograms of the partial width ratio for $h \to \gamma \gamma$ for the subset of neutralino (top), gravitino (lower left) and low-FT models (lower right) that predict $m_h=126\pm 3$ GeV. 
The blue (red) histogram represents models before any ATLAS searches are applied (after the 7 and 8 TeV SUSY searches) while the green (purple) histograms show models that are expected to  
survive the zero-lepton jets plus MET search at 14 TeV, assuming a luminosity of 300 (3000) fb$^{-1}$. The vertical lines show the expected future limits on $r_{\gamma \gamma}$, and are discussed in Section~\ref{sec:step3}.}
\label{figB}
\end{figure}
\begin{figure}[htbp]
\centerline{\includegraphics[width=5.5in]{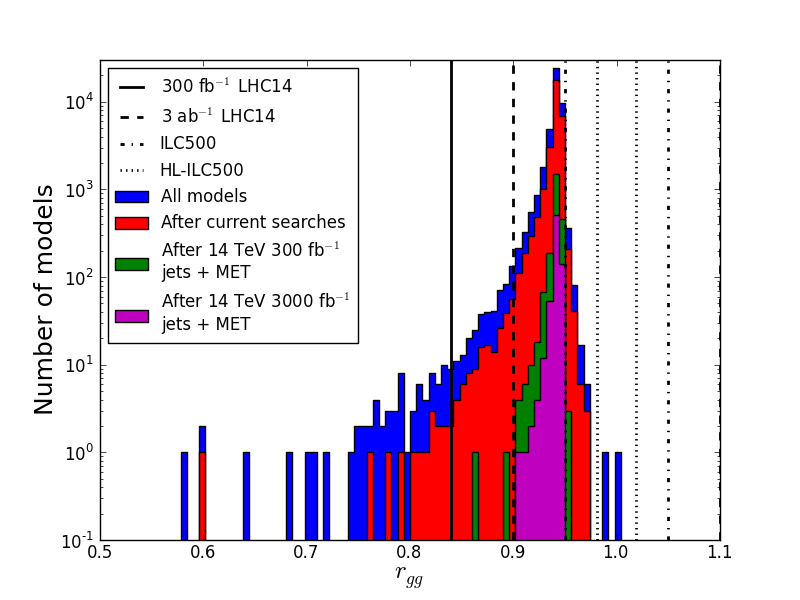}}
\vspace*{0.50cm}
\centerline{\includegraphics[width=3.5in]{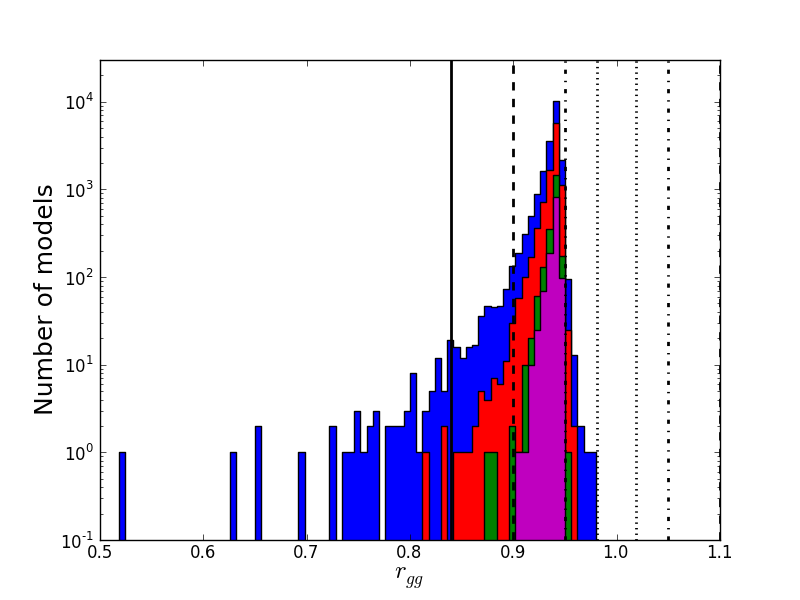}
\hspace{-0.50cm}
\includegraphics[width=3.5in]{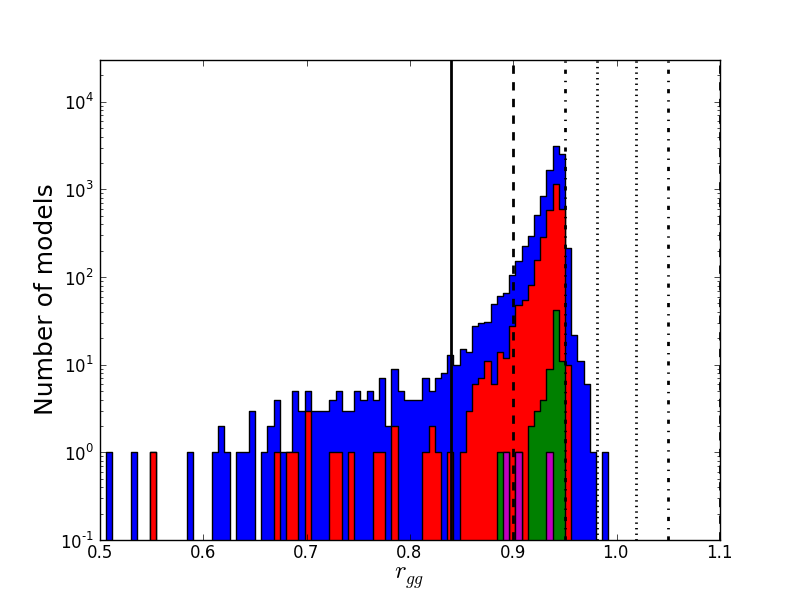}}
\vspace*{-0.10cm}
\caption{Same as the previous Figure but now for $h\to gg$.}
\label{figC}
\end{figure}
\begin{figure}[htbp]
\centerline{\includegraphics[width=5.5in]{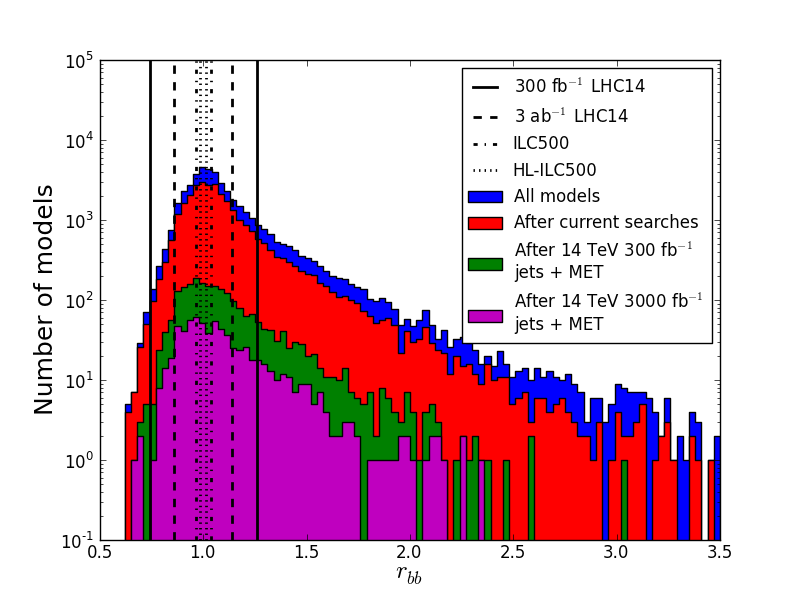}}
\vspace*{0.50cm}
\centerline{\includegraphics[width=3.5in]{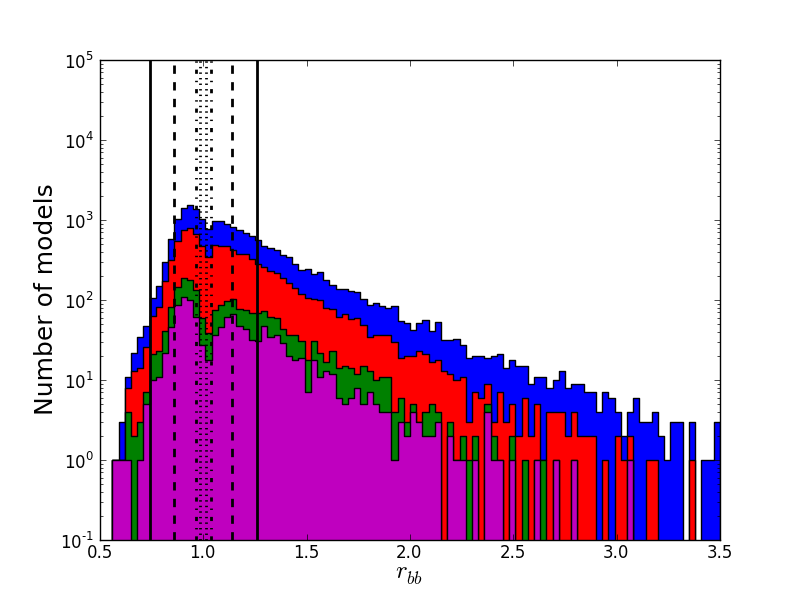}
\hspace{-0.50cm}
\includegraphics[width=3.5in]{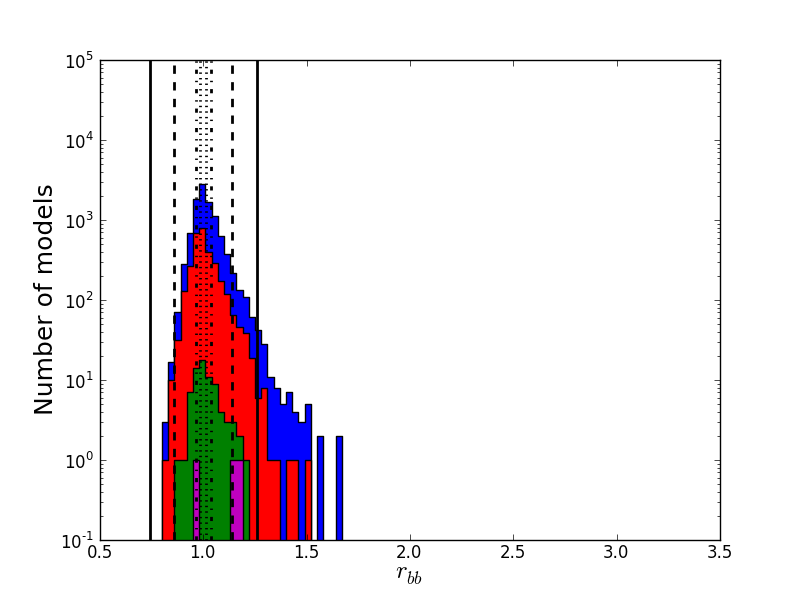}}
\vspace*{-0.10cm}
\caption{Same as in Figure~\ref{figB} but now for $h\to b \bar b$.}
\label{figD}
\end{figure}
\begin{figure}[htbp]
\centerline{\includegraphics[width=5.5in]{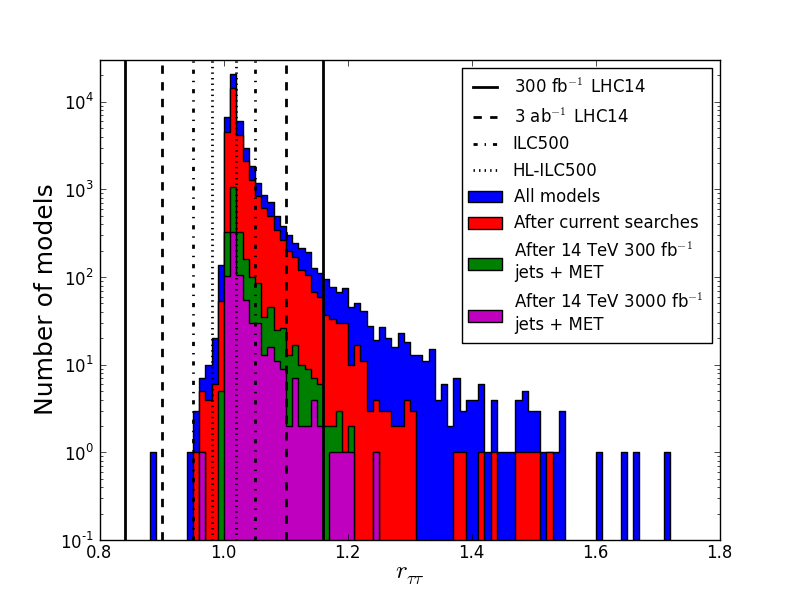}}
\vspace*{0.50cm}
\centerline{\includegraphics[width=3.5in]{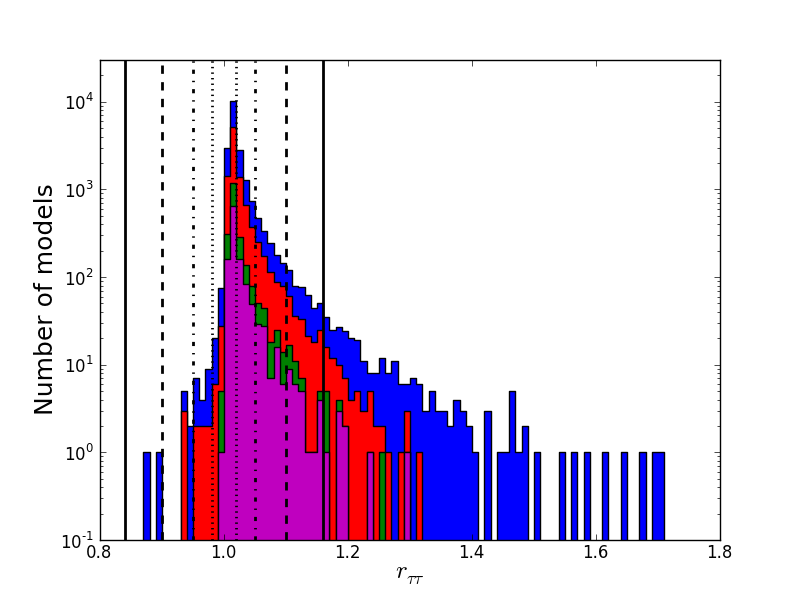}
\hspace{-0.50cm}
\includegraphics[width=3.5in]{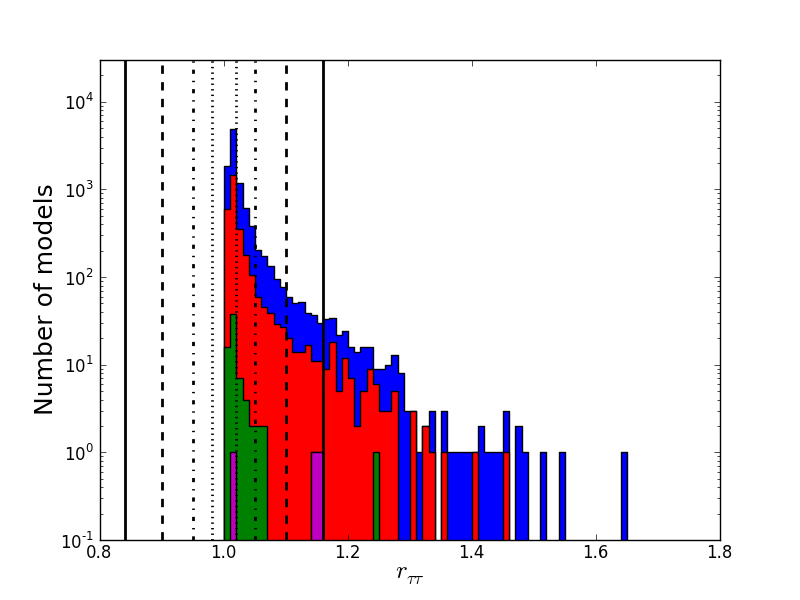}}
\vspace*{-0.10cm}
\caption{Same as in Figure~\ref{figB} but now for $h\to \tau^+\tau^-$.}
\label{figE}
\end{figure}

We now turn our attention to the predicted distributions for the values of the various partial width distributions, $r_X$, in each model set, and the effect of future LHC searches on these distributions. We will return to these distributions in our subsequent analysis to understand the effects of future Higgs coupling measurements. 

Figure~\ref{figB} shows the histogram of values for the SM-normalized partial decay width $r_{\gamma\gamma}$ for the three different model sets. The 
vertical lines appearing in these plots are discussed in the next section. Qualitatively we see that the effect of LHC searches on the $r_{\gamma\gamma}$ distributions is to decrease the normalization while preserving the shapes of the distributions; deviations from this behavior are seen mainly in the tails of the distributions, where the statistics are low. The different responses of each model set to the direct LHC SUSY searches can be seen by observing the widely differing impacts of the searches on the distribution areas. Interestingly, the shape of the distribution of $r_{\gamma\gamma}$ in the neutralino model set is very different from the corresponding distribution of diphoton signal strengths shown in Figure~\ref{figA}, a difference which results mainly from large corrections to the $h \to b \bar{b}$ partial width (which will be discussed below), and therefore to the total width. These corrections alter the diphoton branching fraction, and therefore the signal strength, for a given value of $r_{\gamma\gamma}$. Note also that the distributions of $r_{\gamma\gamma}$ in the neutralino and gravitino model sets are rather similar and somewhat distinct from the corresponding distribution in the low-FT model set, which exhibits a somewhat broader range of values for $r_{\gamma\gamma}$ despite lower statistics. The larger spread in the low-FT distribution arises from the mandatory presence of light charginos, stops, and (in many cases) sbottoms, typically resulting in larger SUSY corrections to the effective $h \gamma\gamma$ coupling than in the neutralino and gravitino model sets, in which charged sparticles are not required to be relatively light. Finally, note that in all three model  
sets the value of $r_{\gamma\gamma}$ peaks at the same value, slightly above unity. We will see below that this shift is reasonably correlated with an opposite shift in the 
peak of the $r_{gg}$ distribution, and that both offsets result from the large stop mixing that is necessary to obtain the correct Higgs mass.

Figure~\ref{figC} displays the corresponding histograms for the values of $r_{gg}$, again showing the distribution for each model set. Once again, the neutralino and gravitino distributions are quite similar while the low-FT distribution is different as a result of distinct requirements on the sparticle spectra. As shown in, \eg, \cite{Carena:2013iba}, the large Higgs mass in the pMSSM generally requires large stop mixing, which results in a small ($\sim 7\%$) but important reduction in the $h \to gg$ partial width and a simultaneous enhancement in the $h \to \gamma\gamma$ partial width. If the stop sector is totally responsible for this shift (which is a reasonable approximation in many cases), then the shift in $r_{gg}$
at the amplitude level is $\sim 3$ times larger than the corresponding shift in $r_{\gamma\gamma}$, with the two shifts having opposite signs. As a result of this effect, nearly all of our models predict $r_{gg}$ to be below unity, an observation which will figure prominently in our subsequent discussion of future experimental constraints on the Higgs couplings. Interestingly, we also see that the tails of the $r_{gg}$ distribution, though present, are not very large. They are slightly larger in the low-FT $r_{gg}$ distribution, since the relevant corrections tend to be larger as a result of the bias towards light stops in the low-FT model set.

Figure~\ref{figD} shows the results for the ratio $r_{bb}$ for the three different model sets, with the neutralino and gravitino distributions again differing somewhat from the low-FT distribution. Small differences between the neutralino and gravitino distributions arise from, \eg, the fact that lighter stops can appear in the gravitino set, since the requirement for the stop to be heavier than the LSP is trivially satisfied by $m_{LSP} \sim 0$ in most of the gravitino LSP models. For each model set we see the now familiar pattern in which the LHC searches do not significantly alter the shapes of the partial width distributions. Unlike the previous cases, however, we now see that $r_{bb}$ may deviate from unity by an $O(1)$ factor. These deviations result from large sbottom mixings that can make $O(1)$ changes in the $hb\bar b$ couplings through non-decoupling 
(mostly gluino) loop effects. These loop effects are driven by the size of the off-diagonal piece of the sbottom mass matrix $m_b(A_b-\mu \tan \beta)$, which is enhanced for large values of $\tan \beta$. While the tail mostly extends to larger values 
of $r_{bb}$, we see that models also exist with $r_{bb}$ significantly below unity. Since the $b\bar b$ mode dominates the Higgs width, this same effect also explains the large spread in the  distribution of $r_{total}$, shown for the neutralino model set in Fig.~\ref{figA} and described above. In our neutralino and gravitino parameter scans, $|A_b|$ and $|\mu|$ are typically of a similar size while $\tan \beta$ has typical values that are $\mathcal{O}(10)$, so it is the $\mu \tan \beta$ term in the off-diagonal piece that dominates. However, in the low-FT set this is no longer true as 
$|\mu|$ is now forced to be relatively small. Thus in the low-FT case we do not expect the range of values for $r_{bb}$ to be as large as it is in the neutralino and 
gravitino sets; this is exactly what we see in Fig.~\ref{figD}.

Figure~\ref{figE} shows the analogous results for the ratio $r_{\tau\tau}$ for the three different model sets. Here we again see that the shapes of the $r_{\tau\tau}$ histograms 
are not significantly altered by the ATLAS SUSY searches at this level of statistics. We also see that the peak occurs at a value slightly greater than unity (by $\sim 2\%$) with a significant tail extending to larger values. This is not surprising since there are also non-decoupling effects in the corrections to the $h\tau \tau$ vertex, although they occur 
via electroweakino loops and are proportional to the $\tau$ mass. This implies that the effect of these non-decoupling terms should be relatively small when compared with their effect in the case of $r_{bb}$, and that is indeed what we observe. Again, since this non-decoupling occurs via the off-diagonal 
$m_\tau (A_\tau -\mu \tan \beta)$ term in the stau mass matrix, these effects should be somewhat suppressed in the low-FT model set in comparison to the other model sets, and this is demonstrated 
in Figure~\ref{figE}.

\begin{figure}[htbp]
\centerline{\includegraphics[width=4.5in]{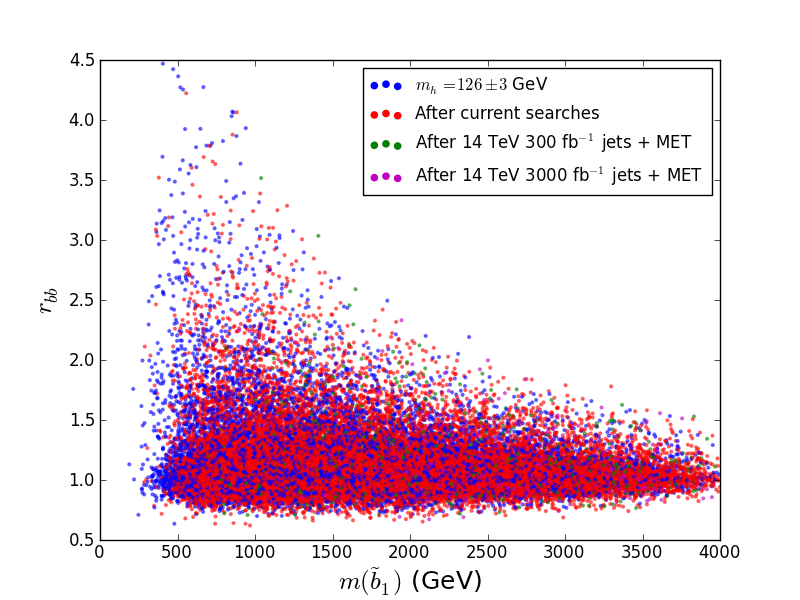}}
\vspace{-0.10cm}
\centerline{\includegraphics[width=4.5in]{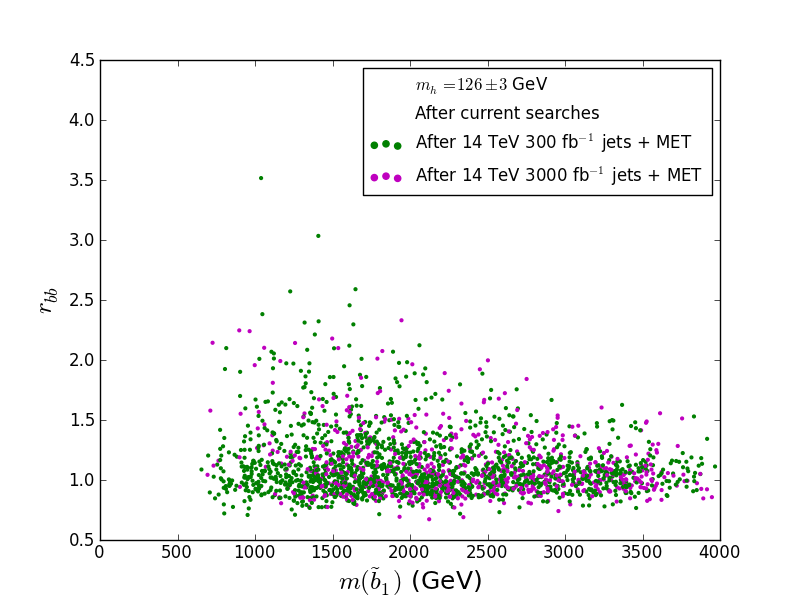}}
\vspace*{-0.10cm}
\caption{Values of $r_{bb}$ as a function of the lightest sbottom mass for the neutralino model set showing the influence of the ATLAS SUSY searches. The lower panel shows those models 
surviving the zero lepton, jets plus MET search at 14 TeV. }
\label{figF}
\end{figure}

Figure~\ref{figF} shows the dependence of $r_{bb}$ on the lighter sbottom mass for the neutralino LSP models as the effects of the direct LHC SUSY searches are imposed. Clearly, measuring a value of this ratio near unity will not impose a constraint on the sbottom mass, regardless of the precision of the measurement. On the other hand, very large deviations of this ratio from unity are seen to require a relatively light sbottom mass, meaning that null SUSY search results should be able to reduce the allowed range for $r_{bb}$. However, the non-decoupling nature of the corrections means that values of $r_{bb}$ above 2 remain allowed even after the 14 TeV jets + MET search is included; excluding O(1) deviations from $r_{bb} = 1$ (which can occur for sbottoms as heavy as 2.5 TeV) through SUSY searches is clearly beyond the capability of the LHC. The large sbottom mass reach necessary to constrain $r_{bb}$ significantly explains our earlier observation that its distribution is roughly independent of applying the LHC searches, although we see now that the 14 TeV searches do begin to have an impact at the edges of the $r_{bb}$ distribution. Although we do not show them here, the corresponding results for the gravitino set are found to be qualitatively similar to those for the neutralino set, although they differ in detail due to the contrasting reach of direct sparticle searches in these two sets. 
Figure~\ref{figFP} shows the analogous result for the low-FT model set. The essential reason for the shapes in this figure  
has already been discussed above; since in the low-FT set $|\mu|$ needs to be relatively small, the size of the off-diagonal element in the sbottom mass matrix is decreased, and the corrections to $r_{bb}$ become significantly smaller.

\begin{figure}[htbp]
\centerline{\includegraphics[width=4.5in]{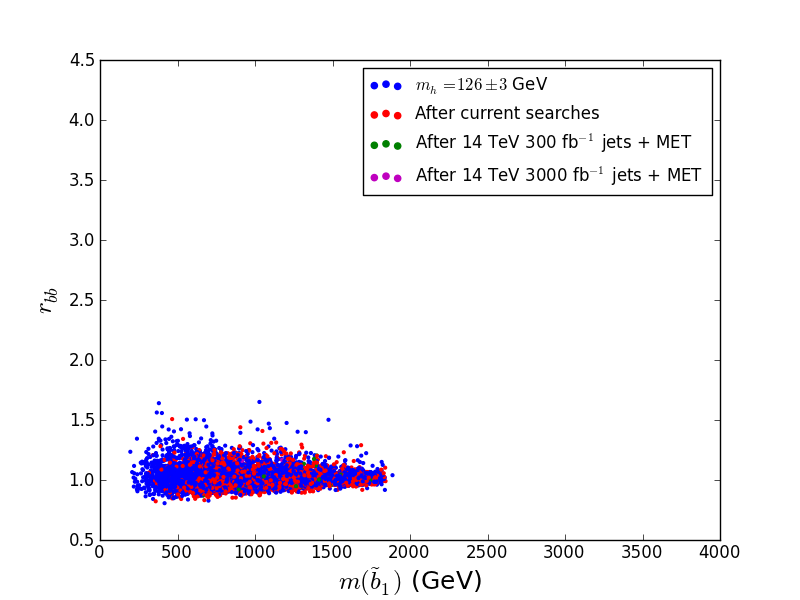}}
\vspace{-0.10cm}
\centerline{\includegraphics[width=4.5in]{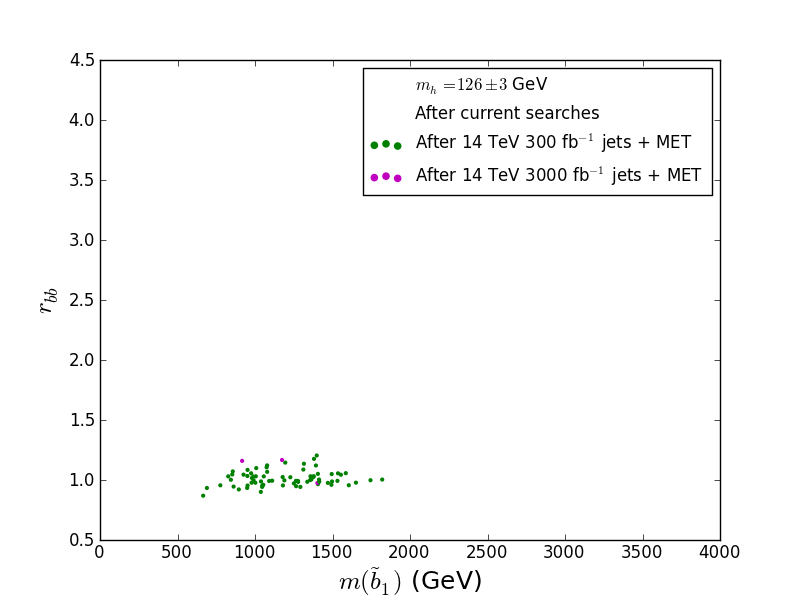}}
\vspace*{-0.10cm}
\caption{Same as the previous Figure but now for the low-FT model set.}
\label{figFP}
\end{figure}

The top panel in Fig.~\ref{figG} shows the Higgs invisible branching fraction, $B(h\to \chi\chi)$, as a function of the LSP mass for the few neutralino LSP models for which 
this is kinematically allowed and how they respond to the direct SUSY searches at the LHC. Note that all of these models have $B(h\to \chi\chi) <0.5$, meaning that they remain unconstrained by current LHC Higgs data. However, we note that all of these models \textit{will} eventually be excluded (or discovered) by sparticle searches as well as by searches for Higgs $\to$ invisible 
at the 14 TeV LHC and/or ILC500. The lower left panel shows the corresponding results for the gravitino set; here we see that for these models a much smaller 
branching fraction is obtained. Of course for these gravitino models the lightest neutralinos will only produce an invisible final state if they escape the detector before decaying; neutralinos with $c \tau \lesssim 1$m will have visible decays, generally producing a (possibly displaced) diphoton + MET signature, where the diphotons would of course fail to reconstruct the Higgs mass. However, the stability of the neutralino tends to be unimportant, since (with the possible exception of the model with the lightest neutralino) the $h\to \chi \chi$ branching fractions seen here are far too small to be accessible at the 14 TeV LHC. The bottom right panel shows the same distribution, now for the low-FT model set. Here we see that the additional constraints imposed on the pMSSM spectrum yield many light 
LSPs which are mainly bino-Higgsino admixtures, a sizeable fraction of which pair-annihilate via the Higgs funnel. In all cases, however, the invisible branching fraction is 
found to be below $\sim 20\%$, which will barely be accessible at the 14 TeV LHC. While many of these models are now excluded by LHC SUSY searches, the remainder would be excluded by 
the 14 TeV jets plus MET search.

\begin{figure}[htbp]
\centerline{\includegraphics[width=5.5in]{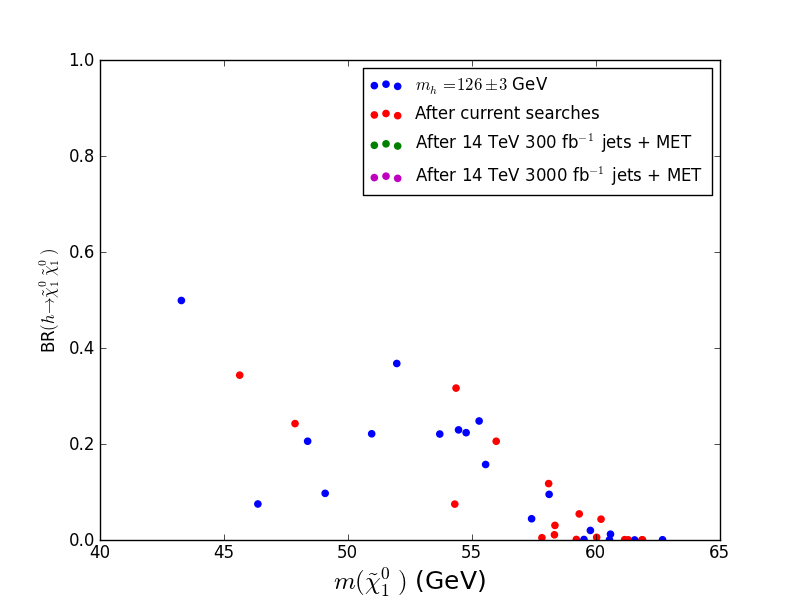}}
\vspace*{0.50cm}
\centerline{\includegraphics[width=3.5in]{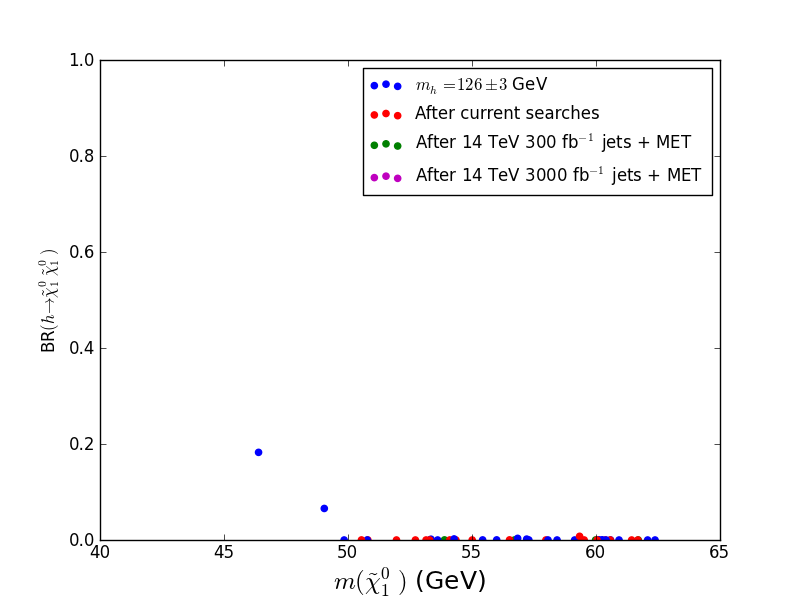}
\hspace{-0.50cm}
\includegraphics[width=3.5in]{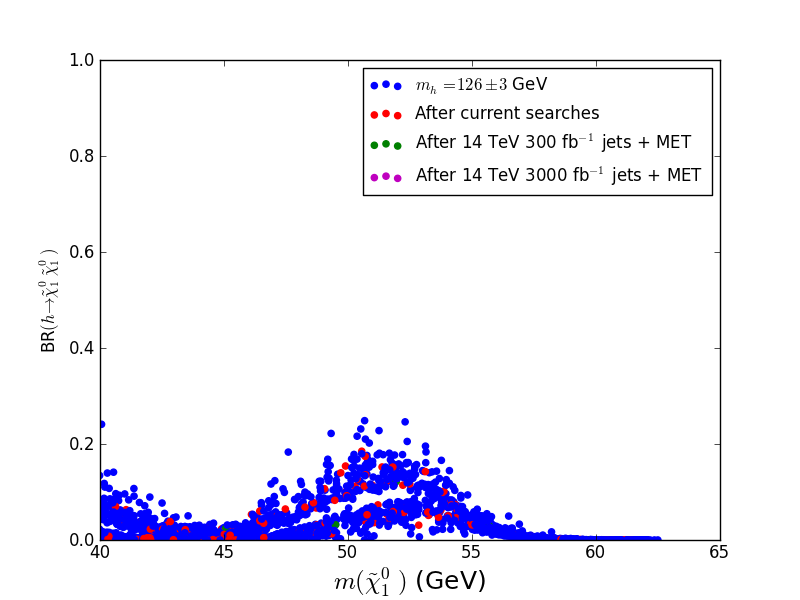}}
\vspace*{-0.10cm}
\caption{(Top) Branching fraction for invisible Higgs decays ($ h \to \chi_1^0 \chi_1^0$) for neutralino LSP models with the correct Higgs mass. The points are 
color-coded according to their response to the LHC SUSY searches. The analogous results for the gravitino (bottom left) and low-FT (bottom right) model sets are also shown.}
\label{figG}
\end{figure}

Although we do not include the ratio $r_{tt}$ directly in our analysis, the predicted values of this quantity found in the various model sets are of some interest. Figure~\ref{figH} displays the 
ratio $r_{tt}$ for the various model sets as a function of the lightest stop mass, where we here have defined $r_{tt}$ as the ratio of the square of the effective 
$ht\bar{t}$ coupling to its SM value, employing the approximation provided in \cite{Djouadi:2013uqa}. Here we see that, \eg, in the case of the low-FT set, significant 
deviations of this ratio away from unity (\eg, $r_{tt}>2$, say) require stop masses below $\simeq 650$ GeV. However, as in the case of the ratio $r_{bb}$ discussed above, a 
measurement of $r_{tt}$ near unity will not exclude any particular range of stop masses in this model set. Note that both the neutralino and gravitino model sets, for which similar results are obtained, can easily produce significant departures of $r_{tt}$ from unity even with relatively heavy stop masses. For these model sets, values of $r_{tt}$ as 
large as 2 can be achieved for lightest stop masses up to $\sim 1.6$ TeV. This wider range of allowed values in the neutralino and gravitino model sets should not be too 
surprising since the stop mass matrix and the resulting physical mass spectrum are not constrained in these models by the additional requirements placed on the low-FT set. 

\begin{figure}[htbp]
\centerline{\includegraphics[width=3.5in]{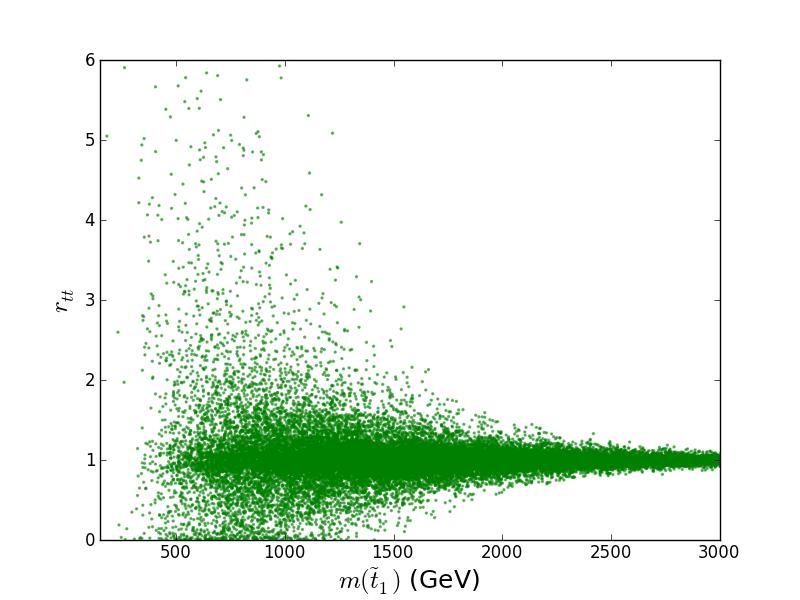}}
\vspace*{0.50cm}
\centerline{\includegraphics[width=3.5in]{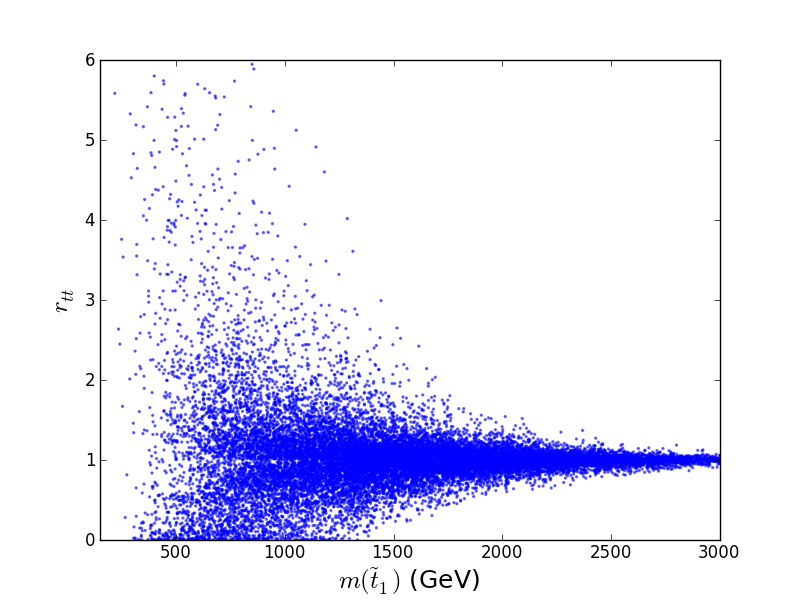}
\hspace{-0.50cm}
\includegraphics[width=3.5in]{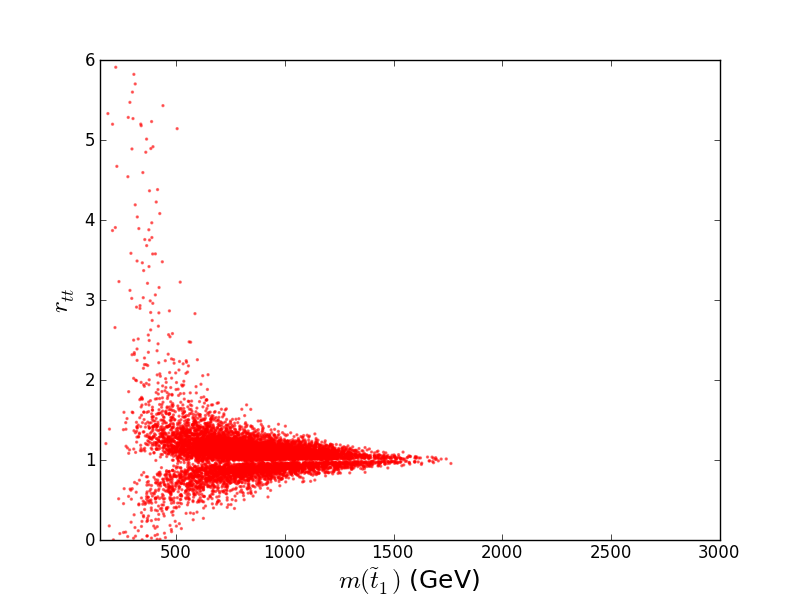}}
\vspace*{-0.10cm}
\caption{The predicted values of the ratio $r_{tt}$ (the squared $ht\bar t$ coupling normalized to its SM value) for the various pMSSM models as a function of the lightest stop mass.
The top (lower left, lower right) panel shows the results for the neutralino (gravitino, low-ft) model set. }
\label{figH}
\end{figure}

\section{Analysis and Results} 
\label{sec:step3}

Now that we have assembled the necessary pieces for our analysis, we can ask how the measurements of the various Higgs couplings at the LHC and ILC will constrain 
the pMSSM parameter space, and how these constraints compare with those from direct SUSY searches. In what follows, we will make use of the numerical results for current and 
future Higgs coupling measurements as presented in Refs.~\cite{now,14TeVLHC,ILC}. Note that in the results quoted below, in taking allowed ranges for the ratios of pMSSM to SM 
Higgs couplings, we will ignore potentially significant theoretical uncertainties (as important higher order corrections have yet to be
calculated) which may be quite important given the claimed level of precision for future LHC and ILC 
coupling measurements. We remind the reader to keep this important issue in mind when interpreting our results; our 
results should thus be treated as indicative only. Clearly more theoretical work will be necessary before (sub-)percent-level measurements are truly meaningful.

We first note that the current LHC data does not significantly constrain the pMSSM parameter space, since the precision of the Higgs coupling measurements is 
still rather low in comparison to the deviations that we expect in our pMSSM model sets. Once 14 TeV LHC and ILC data is available, this will no longer be the case. Even more 
importantly, we note that in order to obtain the results we show below, {\it we have assumed} that the central values of the couplings measured at both the LHC and ILC will 
coincide {\it exactly} with the SM values, and furthermore ignore the theoretical uncertainties associated with the pMSSM predictions themselves  (relative to those of the SM 
at the very least). As the reader will see from the discussion and the Tables that follow, the observation of central values differing from the SM prediction (even within the 
expected ranges) will most certainly exclude a different fraction of our models. This is particularly true for both the loop-sensitive ratios $r_{gg}$ and $r_{\gamma\gamma}$, for 
which the predicted deviations from the SM values in the pMSSM are essentially all in one direction. Of course our qualitative results, which indicate that precise Higgs coupling 
measurements (when properly understood) have significant sensitivity to our pMSSM models, do 
{\it not} depend on the actual central values that will be observed for these 
couplings.  

Assuming that the measured central value of each parameter is equal to the SM prediction, we return to Figs.~\ref{figB},~\ref{figC},~\ref{figD} and~\ref{figE} and now concentrate 
on the vertical lines, which show the expected sensitivity at future experiments. 
These show the regions of the various $r_X$ for the the three model sets which will be allowed or excluded at the $95\%$ CL by Higgs coupling measurements 
at the LHC and HL-LHC{\cite {14TeVLHC}}, the 500 GeV ILC (ILC500) and the ILC500 with a luminosity upgrade{\cite {ILC}}, here denoted as HL-ILC500.  Of course, it is important to 
note that we can always slide these `allowed' regions around to estimate the implications of other possible central value measurements. However, the key result here is that, 
regardless of what central values are actually observed, {\it indirect Higgs coupling measurements will likely result in the exclusion 
(or discovery) of pMSSM models which are not accessible to 
direct SUSY searches at the LHC}. An important caveat to this, of course, is that we need to include (many) more 14 TeV SUSY searches before this result can be shown to be truly robust. However, 
we know from our 7 and 8 TeV studies that the zero-lepton, jets plus MET search will almost certainly be the most powerful search at 14 TeV, as least for the neutralino and low-FT 
model sets, and so this qualitative conclusion is unlikely to change.{\footnote {We note in our companion White Paper~\cite{Cahill-Rowley:2013yla} that the 14 TeV jets+MET search has an impact on the gravitino LSP model set which is diminished, but still significant, compared with its impact on the neutralino LSP model set.}} This result is seen to hold for all of the model sets.

Taking these results at face value (but remembering the above caveats), we can extract some relevant numbers directly from these Figures. The first question we can now address 
is what fraction of the presently allowed (\ie, those passing the 7 and 8 TeV ATLAS analyses and predicting $m_h=126\pm 3$ GeV) pMSSM models will be indirectly excluded by future 
measurements of the Higgs couplings by the LHC and ILC for the three different model sets? The second question is how will these results be modified by the 14 TeV LHC SUSY searches?  
The answers to these questions for both the 14 TeV LHC and the ILC can be seen in the set of Tables~\ref{T1},~\ref{T2} and~\ref{T3}. In these Tables we see a number of 
important results: ($i$) at the LHC, constraining the $hb\bar b$ coupling yields the strongest limits on the allowed pMSSM parameter space. This could also be true at the 
ILC depending on the observed central values for the measured couplings. However, {\it if} we assume that the central values exactly correspond to the SM values, we see that the ILC 
determination of the $hgg$ coupling does much of the damage to the remaining parameter pMSSM space. The reason for this is clear: since $r_{gg}$ is forced to be 
less than unity by the stop mixing required to produce the correct value of the Higgs mass, measuring $r_{gg}=1$ with a very small error will exclude essentially all of 
the model sets! If, on the other hand, the central value were measured to be only $\sim 2-3\%$ {\it below} unity, a much smaller percentage of models would then be excluded. 
For example, if the central value of $r_{gg}$ were measured to be 0.97 with the same errors, then we would find that this measurement would only exclude $2.7\%$ of the 
neutralino LSP models at ILC500 and that $hb\bar b$ would remain the dominant constraint on this model set in this case. We therefore see that in this specific case our results are 
sensitive to our assumption that the measured central values will agree with the predicted SM values. In any case, ($ii$) we see that both the LHC and ILC will provide very 
powerful constraints on the pMSSM model space and have the potential to exclude models that would otherwise remain allowed even after the SUSY searches are performed. In particular, the precision available on Higgs couplings at the ILC will deeply probe the pMSSM parameter space.

Tables~\ref{T1}-\ref{T3} also show that ($iii$) although the general shapes of the $r_X$ distributions are somewhat similar they differ in detail in such a way that the pMSSM model sets 
will respond distinctly to the various indirect Higgs coupling measurement constraints. Of course the ILC500 is extremely powerful in all cases.   
The last thing we notice is ($iv$) that the entries in the Tables do not vary greatly as we include more results from future SUSY searches. This is not surprising; in the limit 
that the shapes of the $r_X$ distributions are completely unaffected by the SUSY search results, the Table entries 
should be independent of which LHC searches have been applied. The limited size of our model samples and the small changes in the $r_X$ distribution shapes account for the 
observed variations.

\begin{table}
\begin{center}
\begin{tabular}{|c||c|c|c|c|}
\hline
Channel & 300 fb$^{-1}$ LHC & 3 ab$^{-1}$ LHC & 500 GeV ILC & HL 500 GeV ILC \\
\hline
\hline

$b\bar{b}$ & 16.4 (27.5, 0.5) & 32.4 (48.5, 5.5) & 77.4 (89.0, 49.0) & 90.5 (95.8, 77.4) \\
$\tau \tau$ & 0.7 (0.7, 3.0) & 3.1 (2.7, 5.8) & 11.6 (9.7, 12.3) & 36.8 (34.2, 32.8) \\
$gg$ & 0.05 (0.03, 0.6) & 0.6 (0.6, 3.1) & 99.1 (99.7, 99.7) & 100.0 (100.0, 100.0) \\
$\gamma \gamma$ & 0.03 (0.06, 0.03) & 0.04 (0.07, 0.2) & 0.03 (0.06, 0.03) & 0.2 (0.15, 0.78) \\

Invisible & --- (---, ---) & --- (---, ---) & 0.03 (0.01, 6.50) & 0.04 (0.01, 7.8) \\
\hline
\hline

All & 16.9 (27.9, 3.9) & 33.9 (49.6, 11.3) & 99.7 (99.96, 99.94) & 100.0 (100.0, 100.0) \\
\hline
\end{tabular}
\caption{The fraction in percent of neutralino (gravitino, low-FT) models with the correct Higgs mass remaining after the current 7 and 8 TeV LHC searches that are expected 
to be excluded by future Higgs coupling measurements, {\it assuming} that the SM values for these couplings are obtained. Blank entries indicate values below 0.01\%. }
\label{T1}
\end{center}
\end{table}
\begin{table}
\begin{center}
\begin{tabular}{|c||c|c|c|c|}
\hline
Channel & 300 fb$^{-1}$ LHC & 3 ab$^{-1}$ LHC & 500 GeV ILC & HL 500 GeV ILC \\
\hline
\hline

$b\bar{b}$ & 20.5 (31.1, 0) & 39.4 (52.8, 6.8) & 82.7 (92.9, 51.4) & 93.1 (97.6, 77.0) \\
$\tau \tau$ & 0.5 (0.6, 1.4) & 3.2 (2.5, 4.1) & 12.8 (9.4, 9.5) & 38.8 (32.2, 27.0) \\
$gg$ & 0 (0, 0) & 0.09 (0.2, 2.7) & 99.9 (99.95, 100.0) & 100.0 (100.0, 100.0) \\
$\gamma \gamma$ & 0 (0, 0) & 0 (0, 0) & 0 (0, 0) & 0 (0, 0) \\

Invisible & --- (---, ---) & --- (---, ---) & 0 (0, 14.9) & 0 (0, 20.3) \\
\hline
\hline

All & 20.8 (31.3, 1.4) & 40.8 (53.6, 9.5) & 99.96 (100.0, 100.0) & 100.0 (100.0, 100.0) \\
\hline
\end{tabular}
\caption{Same as Table~\ref{T1} above but now for the subset of models expected to remain after the ATLAS 14 TeV zero lepton jets + MET search with 300 fb$^{-1}$ of data.}
\label{T2}
\end{center}
\end{table}
\begin{table}
\begin{center}
\begin{tabular}{|c||c|c|c|c|}
\hline
Channel & 300 fb$^{-1}$ LHC & 3 ab$^{-1}$ LHC & 500 GeV ILC & HL 500 GeV ILC \\
\hline
\hline
$b\bar{b}$ & 20.1 (30.9, 0) & 39.1 (53.0, 66.7) & 83.2 (94.1, 66.7) & 93.5 (97.9, 100.0) \\
$\tau \tau$ & 0.7 (0.7, 0) & 3.3 (2.8, 66.7) & 14.3 (9.9, 66.7) & 40.9 (33.9, 66.7) \\
$gg$ & 0 (0, 0) & 0 (0, 33.3) & 100.0 (100.0, 100.0) & 100.0 (100.0, 100.0) \\
$\gamma \gamma$ & 0 (0, 0) & 0 (0, 0) & 0 (0, 0) & 0 (0, 0) \\

Invisible & --- (---, ---) & --- (---, ---) &  0 (0, 0) & 0 (0, 0) \\
\hline
\hline
All & 20.4 (31.1, 0) & 40.1 (54.0, 66.7) & 100.0 (100.0, 100.0) & 100.0 (100.0, 100.0) \\
\hline
\end{tabular}
\caption{Same as Table~\ref{T1} above but now for the subset of models expected to remain after the ATLAS 14 TeV zero lepton jets + MET search with 3 ab$^{-1}$ of data.}
\label{T3}
\end{center}
\end{table}

\section{Conclusion}

In this White Paper we have examined SUSY signals and Higgs boson properties in the context of the pMSSM in models with either neutralino or gravitino LSPs and in models with 
low FT. Within this general scenario we then addressed the following questions: 
`What will potentially null searches for SUSY at the LHC tell us about the possible properties of the Higgs boson?' and, conversely, `What do precision 
measurements of the properties of the Higgs tell us about the possible properties of the various superpartners?' We again warn the reader that in obtaining 
the results presented here we have ignored any theoretical errors associated with the calculation of the Higgs coupling ratios as given by HDECAY. Our results 
can be further refined once a better understanding of this uncertainty is provided by future theoretical work. 

We saw in the above discussion that the answer to the first question was rather straightforward: Given an initial distribution of signal strengths $\mu_X$ or branching 
fraction ratios $r_X$ for a specific final state, the LHC direct SUSY searches reduce the size of the distribution but to a very good approximation do not change its {\it shape}. 
This was shown to be true for all three model sets. This implies that to first order the direct (null) SUSY searches at the LHC will not impact the range of possible deviations 
of Higgs branching fractions from their SM values. This is a very powerful result.  

However, we found the answer to the second question to be much more complex and of potentially even greater importance: Precision measurements of Higgs couplings and 
branching fractions can lead to the exclusion of pMSSM models which cannot be probed by the powerful 14 TeV zero lepton, jets plus MET search, even with an 
integrated luminosity of 3 ab$^{-1}$. This is true for all model sets and also true whether or not the precise values of the measured quantities are consistent with the SM 
expectation. Of course, the more precisely the Higgs couplings are measured, the greater the fraction of pMSSM models that can be 
probed. Since the $hb\bar b$ coupling can deviate the furthest from its SM value within the pMSSM framework, measurements of its value generally have 
the greatest impact {\it if} we do not assume that the central values measured for the Higgs couplings are given exactly by their SM values. If this is 
the case, however, then the $hgg$ coupling at the ILC will provide the strongest constraint as this quantity is necessarily shifted in the pMSSM by stop loops with a central value 
crudely determined by the necessity of obtaining the correct Higgs mass. In such a case (or if the observed central values for $r_{gg}$ -- or to a lesser 
extent $r_{\gamma\gamma}$ -- differ from the SM in the opposite direction from the pMSSM prediction), essentially all of the pMSSM parameter space considered here 
would then be excluded.

\section{Acknowledgments}

We wish to thank M.~Spira for answering our many questions and for assistance with the implementation of the latest version of HDECAY. This work was supported by the Department 
of Energy, Contract DE-AC02-76SF00515.


\begin{thebibliography}{99}


\bibitem{SUSYrefs}
For an overview of MSSM physics and phenomenology as well as mSUGRA, see
  M.~Drees, R.~Godbole, P.~Roy,
  Hackensack, USA: World Scientific (2004) 555 p;
  H.~Baer, X.~Tata,
  Cambridge, UK: Univ. Pr. (2006) 537 p;
  S.~P.~Martin,
  In *Kane, G.L. (ed.): Perspectives on supersymmetry II* 1-153
  [hep-ph/9709356].

\bibitem{Cohen:2013kna} 
  T.~Cohen and J.~G.~Wacker,
  arXiv:1305.2914 [hep-ph].

\bibitem{ATLASH}
See, for example, A. Armbruster, ATLAS Collaboration, talk given at {\it Recontres de Moriond: QCD and High Energy Interaction}, LaThuile, 
March 9-16, 2013.


\bibitem{CMSH}
See, for example, N. Wardle, CMS Collaboration, talk given at {\it Recontres de Moriond: QCD and High Energy Interaction}, LaThuile, 
March 9-16, 2013.


\bibitem{Djouadi:1998di} 
  A.~Djouadi {\it et al.}  [MSSM Working Group Collaboration],
  hep-ph/9901246;
  C.~F.~Berger, J.~S.~Gainer, J.~L.~Hewett and T.~G.~Rizzo,
JHEP {\bf 0902} (2009) 023  [arXiv:0812.0980 [hep-ph]].  


\bibitem{us1} 
  M.~W.~Cahill-Rowley, J.~L.~Hewett, S.~Hoeche, A.~Ismail and T.~G.~Rizzo,
  Eur.\ Phys.\ J.\ C {\bf 72}, 2156 (2012)
  [arXiv:1206.4321 [hep-ph]].

\bibitem{us2} 
  M.~W.~Cahill-Rowley, J.~L.~Hewett, A.~Ismail and T.~G.~Rizzo,
  Phys.\ Rev.\ D {\bf 86}, 075015 (2012)
  [arXiv:1206.5800 [hep-ph]],  
  Phys.\ Rev.\ D {\bf 88}, 035002 (2013)
  [arXiv:1211.1981 [hep-ph]],
  arXiv:1211.7106 [hep-ph] and in preparation. 


\bibitem{Komatsu:2010fb} 
  E.~Komatsu {\it et al.}  [WMAP Collaboration],
  Astrophys.\ J.\ Suppl.\  {\bf 192}, 18 (2011)  
  [arXiv:1001.4538 [astro-ph.CO]].


\bibitem{Allanach:2001kg} 
  B.~C.~Allanach,
  Comput.\ Phys.\ Commun.\  {\bf 143}, 305 (2002)
  [hep-ph/0104145].

\bibitem{Djouadi:2002ze} 
  A.~Djouadi, J.~L.~Kneur and G.~Moultaka,
  Comput.\ Phys.\ Commun.\  {\bf 176}, 426 (2007)
  [arXiv:hep-ph/0211331].


\bibitem{Djouadi:2006bz} 
  A.~Djouadi, M.~M.~Muhlleitner and M.~Spira,
  Acta Phys.\ Polon.\ B {\bf 38}, 635 (2007)  
  [hep-ph/0609292].

 \bibitem{HDECAY}
Thanks to M.~Spira, we employed HDECAY 5.11 in the present analysis which can be obtained from http://people.web.psi.ch/spira/hdecay/. For the original 
reference, see  
  A.~Djouadi, J.~Kalinowski and M.~Spira,
  Comput.\ Phys.\ Commun.\  {\bf 108}, 56 (1998)
  [hep-ph/9704448].

\bibitem{Cahill-Rowley:2013yla} 
  M.~Cahill-Rowley, J.~L.~Hewett, A.~Ismail and T.~G.~Rizzo,
  arXiv:1307.8444 [hep-ph];
  M.~Cahill-Rowley, R.~Cotta, A.~Drlica-Wagner, S.~Funk, J.~Hewett, A.~Ismail, T.~Rizzo and M.~Wood,
  arXiv:1305.6921 [hep-ph].
  
\bibitem{Ellis:1986yg} 
  J.~R.~Ellis, K.~Enqvist, D.~V.~Nanopoulos and F.~Zwirner,
  Mod.\ Phys.\ Lett.\ A {\bf 1}, 57 (1986).
  
\bibitem{Barbieri:1987fn} 
  R.~Barbieri and G.~F.~Giudice,
  Nucl.\ Phys.\ B {\bf 306}, 63 (1988).


\bibitem{us}
  J.~A.~Conley, J.~S.~Gainer, J.~L.~Hewett, M.~P.~Le and T.~G.~Rizzo,
Eur.\ Phys.\ J.\ C {\bf 71} (2011) 1697  [arXiv:1009.2539 [hep-ph]] and   
[arXiv:1103.1697 [hep-ph]].  

\bibitem{CMSextra}
  S.~Chatrchyan {\it et al.}  [CMS Collaboration],
  Phys.\ Lett.\ B {\bf 713}, 408 (2012)
  [arXiv:1205.0272 [hep-ex]];
  S.~Chatrchyan {\it et al.}  [CMS Collaboration],
  Phys.\ Lett.\ B {\bf 713}, 68 (2012)
  [arXiv:1202.4083 [hep-ex]].


\bibitem{BSMUMU} 
  R.~Aaij {\it et al.}  [LHCb Collaboration],
  arXiv:1307.5024 [hep-ex];
  S.~Chatrchyan {\it et al.}  [CMS Collaboration],
  arXiv:1307.5025 [hep-ex].
The analysis as presented here, however, employs the earlier results as given in 
  R.~Aaij {\it et al.}  [LHCb Collaboration],
  Phys.\ Rev.\ Lett.\  {\bf 110}, 021801 (2013)
  [arXiv:1211.2674 [Unknown]]. 
The numerical impact of these later results on our analysis is very minor.


\bibitem{Sjostrand:2006za} 
  T.~Sjostrand, S.~Mrenna and P.~Z.~Skands,
      JHEP {\bf 0605}, 026 (2006)
        [hep-ph/0603175].

\bibitem{PGS}
  J. Conway, PGS4, Pretty Good detector Simulation, http://www.physics.ucdavis.edu/~conway/research/software/pgs/pgs.html .


\bibitem{Beenakker:1996ch}
  W.~Beenakker, R.~Hopker, M.~Spira and P.~M.~Zerwas,
  Nucl.\ Phys.\  B {\bf 492}, 51 (1997)
  [arXiv:hep-ph/9610490];


\bibitem{now}

The Higgs coupling results from ATLAS based on 7 and 8 TeV data can be found at https://twiki.cern.ch/twiki/bin/view/AtlasPublic/HiggsPublicResults,   
while those from CMS can be found at https://twiki.cern.ch/twiki/bin/view/CMSPublic/PhysicsResultsHIG. 

\bibitem{14TeVLHC}
We employ the results as presented by ATLAS in ATL-PHYS-PUB-2012-004 and 
by CMS as presented by J.~Olsen at the {\it Snowmass Energy Workshop}, U. of Washington, Seattle, 7/1/2013.  In the numerical analysis presented here, we 
make explicit use of these projected CMS Higgs coupling results for the 14 TeV LHC assuming the larger error for both luminosities. 




\bibitem{ILC}
We follow the results as presented by 
  M.~E.~Peskin,
  arXiv:1207.2516 [hep-ph]; 
  M.~Klute, R.~Lafaye, T.~Plehn, M.~Rauch and D.~Zerwas,
  Europhys.\ Lett.\  {\bf 101}, 51001 (2013)
  [arXiv:1301.1322 [hep-ph]];
The ILC Technical Design Report: Vol.2, H. Baer \etal , 2013. 
Here we make explicit use of the Higgs coupling analysis results as presented in `ILC Higgs White Paper', D.~Asner \etal, submitted to Snowmass 2013. 

\bibitem{Carena:2013iba} 
  M.~Carena, S.~Gori, N.~R.~Shah, C.~E.~M.~Wagner and L.~-T.~Wang,
  arXiv:1303.4414 [hep-ph].


\bibitem{Djouadi:2013uqa} 
  A.~Djouadi, L.~Maiani, G.~Moreau, A.~Polosa, J.~Quevillon and V.~Riquer,
  arXiv:1307.5205 [hep-ph].




\end{thebibliography}
\end{document}